\documentclass{article}
\usepackage[preprint]{neurips_2025}
\makeatletter
\renewcommand{\@noticestring}{Preprint. NeurIPS 2025 Workshop on AI for Music.}
\makeatother
\workshoptitle{AI for Music}

\usepackage[utf8]{inputenc}
\usepackage[T1]{fontenc}
\usepackage[colorlinks,citecolor=brown]{hyperref}
\usepackage{url}
\usepackage{booktabs}
\usepackage{amsfonts}
\usepackage{nicefrac}
\usepackage{microtype}
\usepackage{xcolor}
\usepackage{graphicx}
\usepackage{booktabs,multirow,siunitx,xcolor,colortbl}
\usepackage{placeins}

\title{Generating Piano Music with Transformers: A Comparative Study of Scale, Data, and Metrics}

\newcommand{\samethanks}[1][\value{footnote}]{\footnotemark[#1]}

\author{
    Jonathan Lehmkuhl\thanks{Equal contribution. RWTH Aachen University. Email: \texttt{\{jonathan.lehmkuhl, abel.ilyes-kun, nico.bremes, kaan.oezaltan, frederik.muthers\}@rwth-aachen.de}.} \\
    \And
    Ábel Ilyés-Kun\samethanks \\
    \And
    Nico Bremes\samethanks \\
    \And
    Cemhan Kaan Özaltan\samethanks \\
    \And
    Frederik Muthers\samethanks \\
    \And
    Jiayi Yuan\thanks{University of Washington. Email: \texttt{jiayiy9@cs.washington.edu}.} \\
}

\begin{document}

\maketitle

\begin{abstract}
Although a variety of transformers have been proposed for symbolic music generation in recent years, there is still little comprehensive study on how specific design choices affect the quality of the generated music. In this work, we systematically compare different datasets, model architectures, model sizes, and training strategies for the task of symbolic piano music generation. To support model development and evaluation, we examine a range of quantitative metrics and analyze how well they correlate with human judgment collected through listening studies. Our best-performing model, a 950M-parameter transformer trained on 80K MIDI files from diverse genres, produces outputs that are often rated as human-composed in a Turing-style listening survey.
\end{abstract}

\section{Introduction}
\label{sec:intro}
Music exhibits hierarchical structure across multiple timescales, making transformer models with self-attention well-suited for modeling musical dependencies. Recently proposed transformer models can generate expressive sequences that capture both local patterns and global structure of MIDI data \citep{huangmusic,huang2020pop}. A variety of MIDI datasets \citep{hawthorne2018enabling,raffel2016learning,bradshawaria}, tokenization schemes \citep{oore2020time,huang2020pop,von2023figaro}, transformer-based architectures \citep{yu2022museformer,jin2022transformer,guo2025moonbeam}, and evaluation strategies \citep{yang2020evaluation,retkowski2024frechet} have been proposed in prior work on symbolic music generation. However, due to the lack of a unified evaluation framework for generated music, it remains an open challenge to determine how specific design choices in custom architectures, e.g., the tokenizer, embedding function, or attention mechanism, affect the quality of the generated music. While automatically computable metrics such as perplexity and musically informed objective metrics~\cite{yang2020evaluation} are commonly used, their relationship to human perception of musical quality is not yet fully understood. We aim to address these challenges by presenting an empirical study on piano music generation using transformer models.

\section{Methods}
\label{sec:methods}
\paragraph{Models and training setups}
Our methodology is structured around five targeted experiments. First, we train scaled-down versions (62M, 155M, 439M, and 950M) of the Mistral 7B architecture \citep{jiang2023mistral7b} with sliding window attention and rotary position embedding on the MAESTRO dataset \citep{hawthorne2018enabling}. Furthermore, we pre-train models on a subset of the Aria-MIDI dataset \citep{bradshawaria} and compare them to models trained only on MAESTRO. We fine-tune a model pre-trained on Aria-MIDI on MAESTRO and compare it to a model trained from scratch only on MAESTRO to assess the effect of transfer learning.  To guide generation, we prepend genre labels during training and examine their impact. A genre-conditioned 950M-parameter model is further evaluated in a Turing-like listening study. Finally, we fine-tune the Moonbeam foundation model \citep{guo2025moonbeam} on MAESTRO to provide a high-level benchmark for our custom Mistral-based models. Unlike standard lookup-based embeddings, Moonbeam uses a trainable sinusoidal embedding function \citep{guo2023domain} that enforces translational invariance and pitch transposability, claimed to yield more musically meaningful representations and improved test perplexity.

\paragraph{Tokenization and data preprocessing}
For tokenization, we use the REMI scheme \citep{huang2020pop}, a widely adopted method that imposes metrical structure through position and bar tokens. Our REMI vocabulary contains 485 base tokens, which we extend using byte-pair encoding to learn a 30K-token vocabulary for more efficient representation. To balance efficiency and expressiveness, we fix the model context length to 1024 tokens, typically corresponding to 1–2 minutes of music. All MIDI files in our datasets are split into chunks of roughly 1024 tokens, and we generate samples of the same length during inference. Moonbeam compounds multiple events into single note-level tokens, producing sequences three times shorter than REMI. For a fair comparison, we adjust Moonbeam's context length to match the effective amount of training data seen by the Mistral-based models.

\paragraph{Datasets and data augmentation}
Various datasets of piano music in MIDI format are now available for symbolic music research \citep{hawthorne2018enabling,edwards2023pijama,kong2020giantmidi,zhang2022atepp,ferreira_aiide_2020_adl,bradshawaria}. For our experiments, we use the MAESTRO dataset \citep{hawthorne2018enabling}, containing 200 hours of professional classical piano, and a curated subset of 100K files from Aria-MIDI \citep{bradshawaria}, which spans multiple genres of automatically transcribed piano recordings. To improve generalization and robustness, we augment the data by randomly altering pitch, velocity, note duration, and tempo within perceptually plausible ranges.

\paragraph{Subjective evaluation}
Subjective evaluation is our target metric, as our goal is to generate music perceived as enjoyable, original, and human-like. We conduct listening tests in which five participants rate samples on a five-point Likert scale along three dimensions: \textbf{pleasingness} (how enjoyable the music is), \textbf{authenticity} (how natural or human-like it sounds), and \textbf{novelty} (how original or unique it feels) \citep{yang2017midinet}, with samples presented in randomized order to avoid bias. Inspired by the original Turing Test \citep{alan1950a}, we also conduct a musical Turing-like test, asking listeners to classify excerpts as human-composed, AI-generated, or uncertain.

\paragraph{Objective evaluation}
We use \textbf{perplexity} (PPL) to assess token-level fit, though it does not capture higher-level musical structure. To evaluate global patterns, we employ distributional metrics comparing feature-level distributions of generated and training samples. \textbf{Fréchet music distance} (FMD) \citep{retkowski2024frechet} measures distances between multivariate Gaussian embeddings extracted from MIDI files. We also compute mean and standard deviation of musical features (pitch count, range, intervals, note count, inter-onset interval) and compare datasets via histograms of intra-set and inter-set Euclidean distances, smoothed with kernel density estimation. Similarity is quantified using \textbf{Kullback–Leibler divergence} (KLD) \citep{kullback1951information} and \textbf{overlapping area} (OA), with low KLD and high OA indicating strong resemblance to the training data. We compute the above metrics at intermediate checkpoints during training.

\section{Experiments and results}
\label{sec:experiments}
\begin{figure}[ht]
  \centering
  \includegraphics[width=1\textwidth]{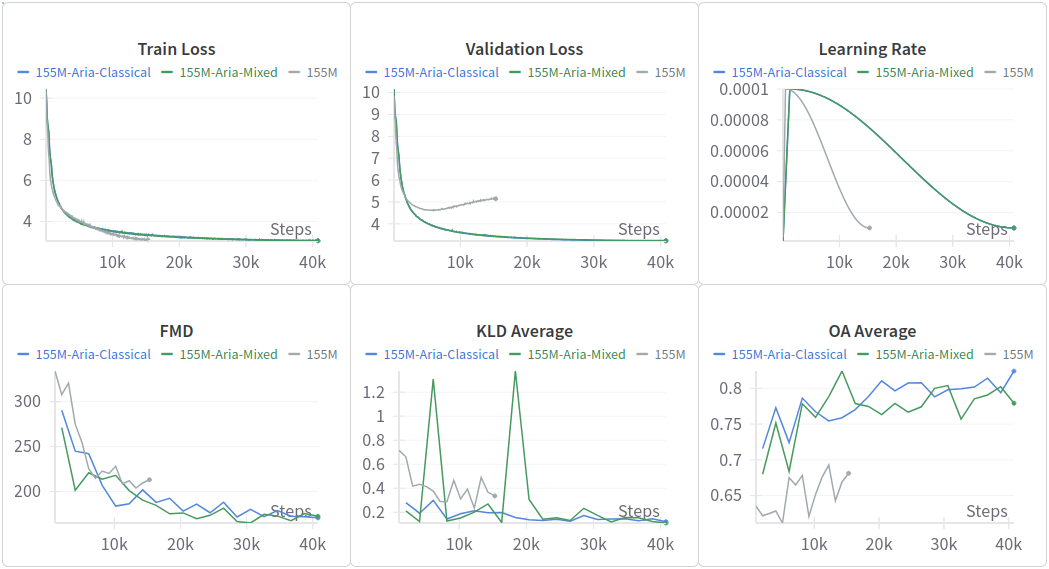}
  \caption{Training curves for the models pre-trained on Aria-Deduped compared to the 155M model trained only on MAESTRO.}
  \label{fig:supp_curve_2}
\end{figure}

\begin{figure}[ht]
  \centering
  \includegraphics[width=1\textwidth]{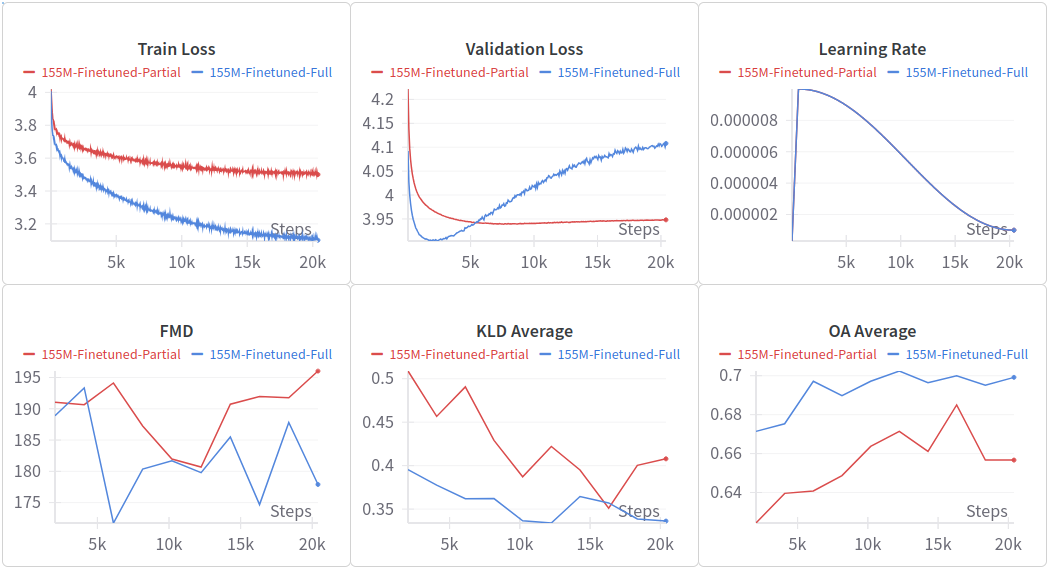}
  \caption{Training curves for the models pre-trained on Aria-Deduped and fine-tuned on MAESTRO.}
  \label{fig:supp_curve_3}
\end{figure}

\begin{figure}[ht]
  \centering
  \includegraphics[width=1\textwidth]{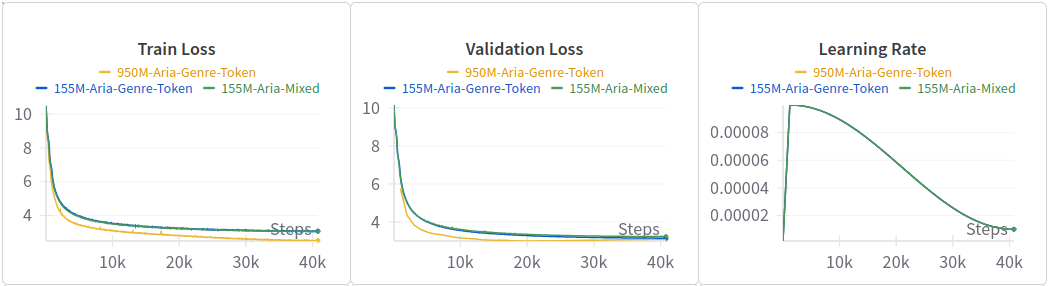}
  \caption{Training curves for the models with integrated genre information.}
  \label{fig:supp_curve_4}
\end{figure}

\begin{figure}[ht]
  \centering
  \includegraphics[width=1\textwidth]{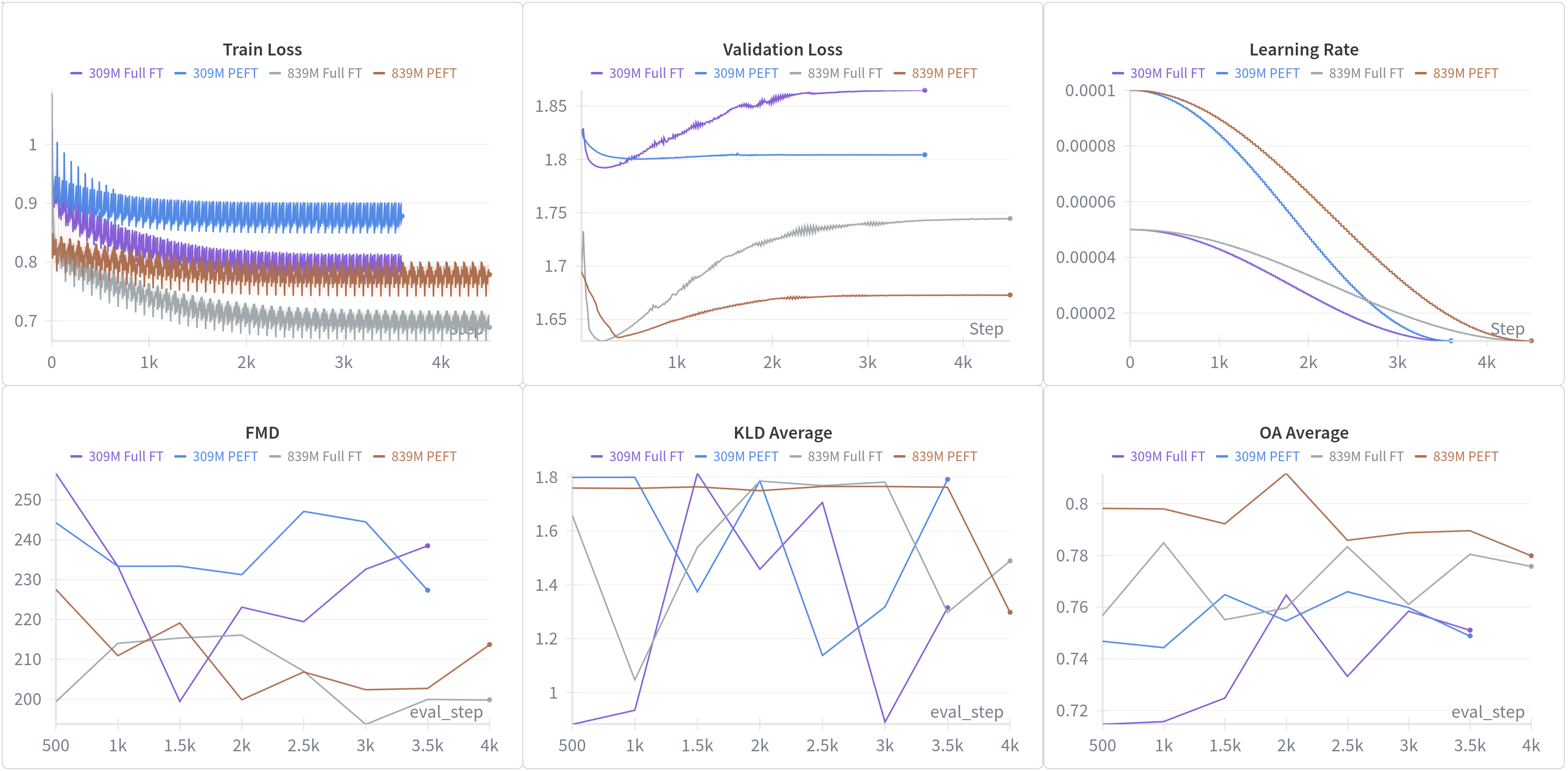}
  \caption{Fine-tuning curves for Moonbeam with context size 512.}
  \label{fig:moonbeam_ctx_512}
\end{figure}

\begin{figure}[ht]
  \centering
  \includegraphics[width=1\textwidth]{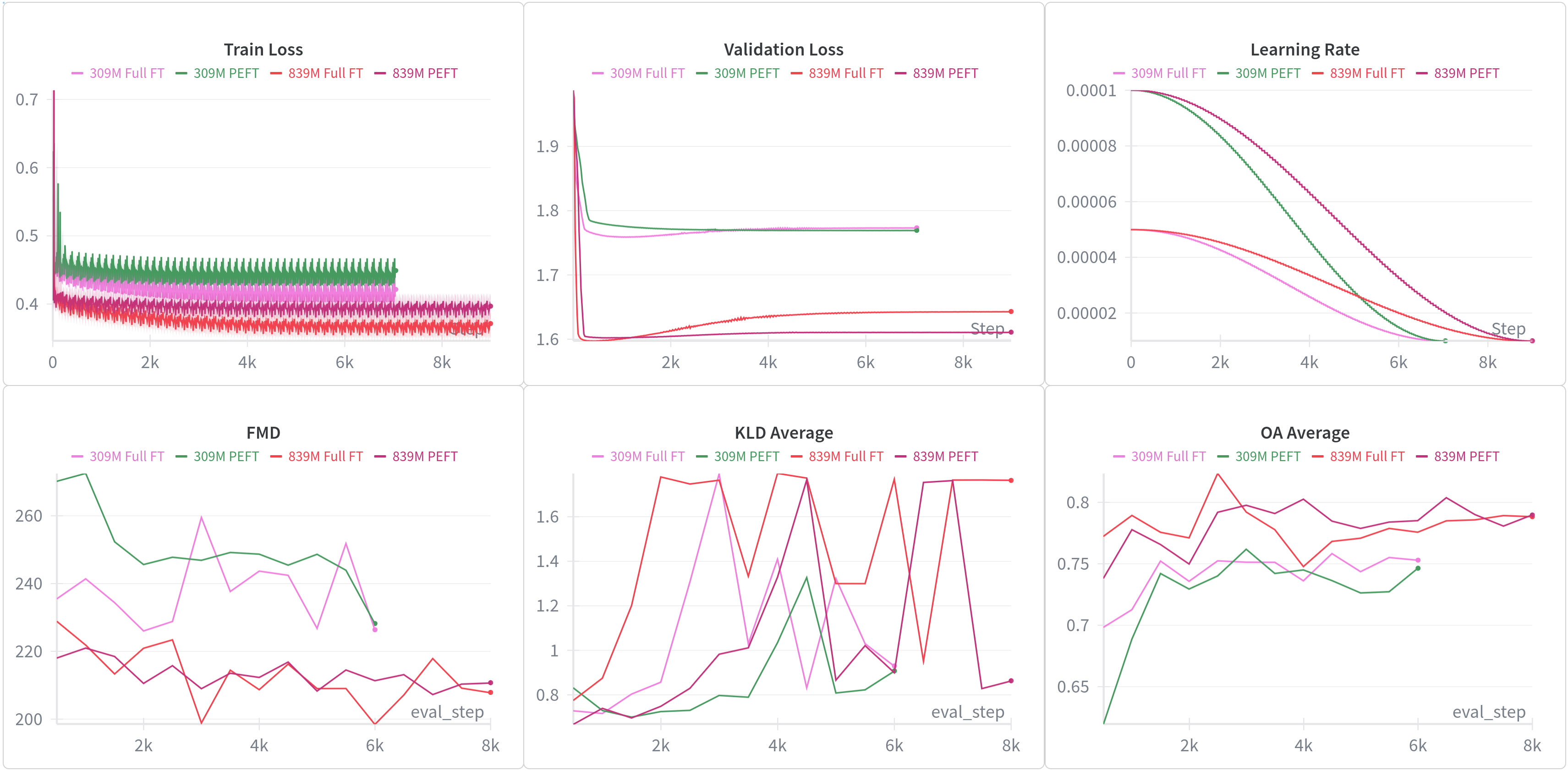}
  \caption{Fine-tuning curves for Moonbeam with context size 1024.}
  \label{fig:moonbeam_ctx_1024}
\end{figure}

\begin{figure}[ht]
  \centering
  \begin{minipage}[b]{0.48\textwidth}
    \centering
    \includegraphics[scale=0.6]{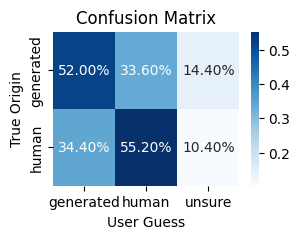}
    \caption{Confusion Matrix for the musical Turing-like test.}
    \label{fig:turing_confusion_matrix}
  \end{minipage}
  \hfill
  \begin{minipage}[b]{0.48\textwidth}
    \centering
    \includegraphics[scale=0.6]{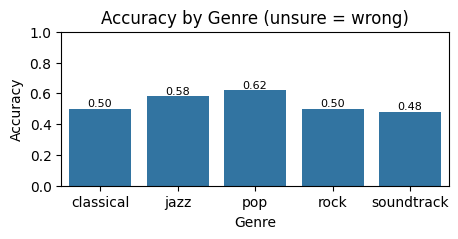}
    \caption{Accuracy by genre in the musical Turing-like test.}
    \label{fig:turing_accuracy_by_genre}
  \end{minipage}
\end{figure}

\begin{figure}[!h]
  \centering
  \includegraphics[scale=0.6]{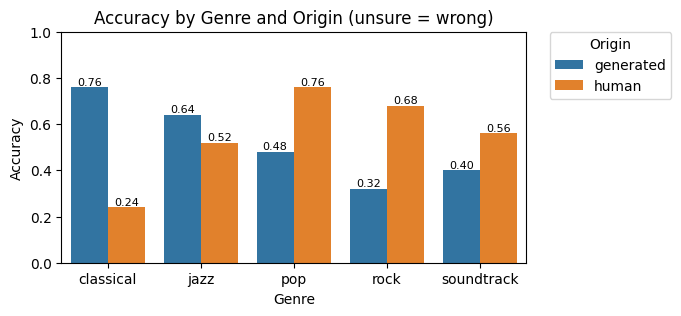}
  \caption{Accuracy by genre and origin in the musical Turing-like test.}
  \label{fig:turing_accuracy_by_genre_origin}
\end{figure}

\begin{figure}[!h]
  \centering
  \includegraphics[scale=0.6]{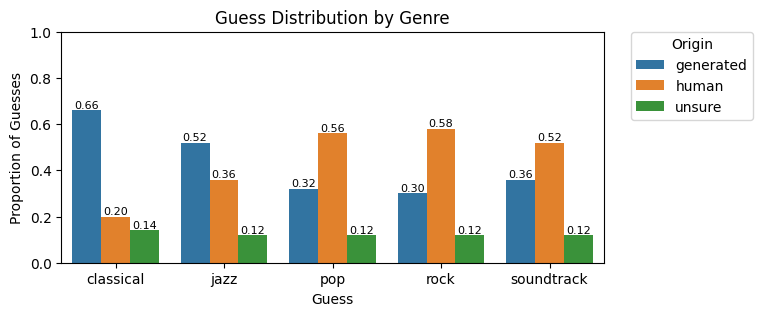}
  \caption{Distribution of participant guesses by genre in the musical Turing-like test.}
  \label{fig:turing_guess_distribution_by_genre}
\end{figure}

\begin{figure}[!h]
  \centering
  \begin{minipage}[b]{0.48\textwidth}
    \centering
    \includegraphics[scale=0.6]{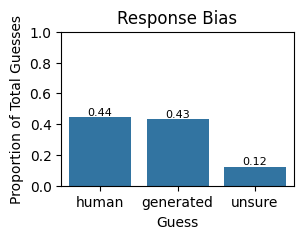}
    \caption{Overall response bias in the musical Turing-like test.}
    \label{fig:turing_response_bias}
  \end{minipage}
  \hfill
  \begin{minipage}[b]{0.48\textwidth}
    \centering
    \includegraphics[scale=0.6]{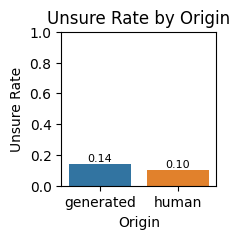}
    \caption{Unsure rate by origin in the musical Turing-like test.}
    \label{fig:turing_unsure_rate_by_origin}
  \end{minipage}
\end{figure}

We provide detailed results as supplementary material in the appendix: Training curves are shown in Figures~\ref{fig:supp_curve_1}--\ref{fig:moonbeam_ctx_1024}, the results of the musical Turing-like test are in Table~\ref{tab:turing_classification_report} and Figures~\ref{fig:turing_confusion_matrix}--\ref{fig:turing_unsure_rate_by_origin}, and Tables~\ref{tab:maestro_full}--\ref{tab:fine-tune_full} present the complete results of the objective and subjective metrics. Our code\footnote{\url{https://github.com/kaanozaltan/piano-transformer}} is available online.

\subsection{Different models on MAESTRO}
We trained four Mistral-based models of varying sizes on the MAESTRO dataset for 225 epochs each. Subjective evaluation of generated samples reveals that, despite overfitting, generation quality generally improves with model size. For overfitting models, later checkpoints often produce more musically coherent outputs, even as validation loss worsens. While this might suggest memorization, our listening impressions indicate otherwise: the samples, though more structured and stylistically consistent, remain too flawed to be mistaken for exact copies of training pieces. This points to a fundamental tradeoff when training large models on small datasets like MAESTRO, especially for creative domains. To learn high-level musical features, the models must overfit to some extent. While this limits generalization, it improves alignment with the training distribution\textemdash{}our primary goal in this setting. Table~\ref{tab:maestro_eval} presents subjective evaluation results alongside PPL, FMD, KLD, and OA for the final checkpoints. FMD, KLD, and OA closely align with human judgments in ranking model quality, whereas PPL diverges for the two largest models\textemdash{}supporting the view that token-level metrics are insufficient for evaluating musical plausibility. Interestingly, the 950M model underperforms the 439M model not only on KLD and OA, but also in the subjective evaluation at the final checkpoint. 

For fine-tuning, we used the two available Moonbeam model checkpoints with 309M and 839M parameters, both pre-trained on a collection of datasets worth over 80K hours of music. We fine-tuned both models for 150~epochs on MAESTRO. We selected the 839M fine-tuned model and a context size of 512 for comparison with our custom models. We calculated the objective metrics on the same reference set as for our custom models with a fidelity of 1000 samples at the checkpoint corresponding to step 3000. Table~\ref{tab:maestro_eval} shows the comparison of the objective metrics and the scores of the subjective listening test across the models. The 439M and 950M custom models outperform Moonbeam on the subjective listening test.

\begin{table}[ht]
  \caption{Subjective and objective evaluation of the custom models and the Moonbeam model trained on MAESTRO. Mean S. column indicates mean subjective evaluation scores. MB indicates the Moonbeam model. PPL not computed for Moonbeam.}
  \vspace{1em}
  \label{tab:maestro_eval}
  \centering
    \begin{tabular}{lcccccccc}
      \toprule
      Model & Mean S. & Pleas. & Auth. & Nov. & FMD\,$\downarrow$ & KLD\,$\downarrow$ & OA\,$\uparrow$ & PPL\,$\downarrow$ \\
      \midrule
      62M  & 2.84 & 2.85 & 2.78 & 2.89 & 210.01 & 0.36 & 62.85\% & 53.65\\
      155M & 2.86 & 2.89 & 2.66 & 3.03 & 187.53 & 0.38 & 68.46\% & 21.32\\
      439M & \textbf{3.22} & \textbf{3.30} & \textbf{3.02} & \textbf{3.35} & \textbf{176.40} & \textbf{0.25} & 71.60\% & 4.82\\
      950M & 3.17 & 3.27 & 2.98 & 3.27 & 176.80 & 0.29 & 68.33\% & \textbf{2.78}\\
      MB 839M & 2.91 & 2.90 & 2.74 & 3.08 & 194.70 & 0.51 & \textbf{82.21}\% & --\\
      \bottomrule
    \end{tabular}
\end{table}

\subsection{Fine-tuning on MAESTRO}
While the Aria dataset provides a large and diverse collection of MIDI files, it is automatically transcribed from internet audio, which can introduce noise and errors. MAESTRO, in contrast, contains high-quality recordings by professional pianists, offering more reliable musical detail. To leverage Aria's scale while benefiting from MAESTRO's quality, we applied transfer learning by fine-tuning models pre-trained on the multi-genre subset of Aria-Deduped. We compared partial fine-tuning (155M-F-P), freezing the first eight blocks to retain general musical structure, with full fine-tuning (155M-F-F), updating all parameters to fully adapt to MAESTRO.

As shown in Table~\ref{tab:fine-tuning_eval}, both variants outperform the MAESTRO-only baseline across most metrics. Full fine-tuning achieves the best overall performance, suggesting that allowing all layers to adapt is beneficial, likely due to the genre gap between the multi-genre pre-training data and classical MAESTRO. The underperformance of partial fine-tuning indicates that low-level musical representations differ across genres and benefit from complete adaptation.

\begin{table}[ht]
  \centering
  \caption{Subjective and objective evaluation of models pre-trained on Aria-Deduped and fine-tuned on MAESTRO compared with the 155M MAESTRO model. Mean S. column indicates mean subjective evaluation scores.
  }
  \vspace{1em}
  \label{tab:fine-tuning_eval}
    \begin{tabular}{lccccccc}
      \toprule
      Model & Mean S. & Pleas. & Auth. & Nov. & FMD\,$\downarrow$ & KLD\,$\downarrow$ & OA\,$\uparrow$ \\
      \midrule
      155M     & 2.85 & 2.80 & 2.80 & 2.95 & 187.53 & 0.38 & 68.46\% \\
      155M-F-P & 3.09 & 3.13 & 2.95 & 3.20 & 193.45 & 0.34 & 70.93\% \\
      155M-F-F & \textbf{3.25} & \textbf{3.30} & \textbf{3.10} & \textbf{3.35} & \textbf{187.34} & \textbf{0.30} & \textbf{72.39\%}\\
      \bottomrule
    \end{tabular}
\end{table}

\subsection{Integrating genre information}
The Aria dataset contains multiple musical genres with differing distributional properties. To better control output style, we trained a model conditioned on genre tokens. Each file's genre was extracted from the metadata, embedded into the MIDI, and prepended as a token to the sequence. During generation, the model could be conditioned by providing only the genre token. We trained a 155M-parameter model on 35K MIDI files from the same Aria subset used previously, allowing direct comparison with a model trained without genre tokens. Results show only minor differences in objective metrics, suggesting that genre tokens alone do not significantly improve the model's understanding of the data. Subjective listening, however, indicated that genre-conditioned samples align well with the intended style, with the 950M model producing particularly high-quality outputs where conditioned genres were clearly apparent.

\subsection{Musical Turing-like test}
Since the 950M model with genre conditioning produced the highest-quality samples, we conducted a musical Turing-like test using sequences from the five most prevalent Aria-Deduped genres. For each genre, five generated and five human-composed sequences were selected from a pool of thirty samples.
Excluding “unsure” responses, participants achieved 61.2\% accuracy; counting “unsure” as incorrect reduced it to 53.6\%. Precision, recall, and F1-scores were similar for human and generated samples, and the confusion matrix showed roughly a third of samples misclassified, indicating many generated sequences were perceptually similar to human music. Accuracy by genre ranged from 48\% (soundtrack) to 62\% (pop), with biases: classical and jazz leaned “generated,” while pop, rock, and soundtrack leaned “human.” Overall, responses were 44.4\% human, 43.2\% generated, and 12.4\% unsure, showing slight bias. Unsure rates were slightly higher for generated sequences (14.4\% vs. 10.4\%), suggesting confusion remained high across genres.

\section{Conclusion}
\label{sec:conclusion}
This study systematically compared transformer models for unconditional piano music generation, varying model scale, dataset size, pre-training and fine-tuning strategies, conditioning methods, and architectures. Larger models generally improved subjective quality, though excessive overfitting on the small MAESTRO dataset sometimes reduced performance. Objective metrics FMD, KLD, and OA aligned reasonably well with human judgments, whereas PPL consistently improved with model size but did not always reflect perceived quality. Pre-training on the larger Aria-MIDI dataset enhanced generation quality and mitigated overfitting, with fine-tuning on MAESTRO further improving both subjective and objective results. Genre conditioning allowed style-specific control, and the largest conditioned model frequently produced outputs rated as human-composed in a Turing-style listening test.

Training larger models on MAESTRO revealed a trade-off between overfitting and musical coherence. Overfitting sometimes improved alignment with the training distribution, producing more coherent outputs without simple memorization, but excessive overfitting could reduce quality. On larger datasets such as Aria-MIDI, this trade-off was less pronounced. While FMD, KLD, and OA generally correlated with subjective judgments, no metric fully captured musical quality. Dataset scale and transfer learning were critical, with pre-training on Aria-MIDI improving generalization and enabling MAESTRO fine-tuning to surpass MAESTRO-only models. 

While Moonbeam's domain-specific embedding space offers theoretical advantages, our custom models performed competitively despite relying on conventional lookup embeddings. This raises questions about the practical benefits of such specialized architectures and whether specific factors may have limited the effectiveness of fine-tuning Moonbeam on MAESTRO. 

Future work should further explore the trade-off between overfitting and generation quality, particularly using high-level musical metrics rather than cross-entropy loss, to better understand when overfitting begins to harm outputs. The benefits of specialized architectures for music generation remain an open question. Future work could probe the features captured at different layers of the model and assess whether the embedding space preserves pitch-transposition relationships in deeper layers. This could be evaluated by measuring reconstruction loss on transposed inputs at various stages in the network. The establishment of robust benchmarks for fair comparison between models from different studies would be desirable. Standardized evaluation frameworks would greatly improve the comparability and reproducibility of results in this field.

\begin{ack}
Computations were performed with computing resources granted by RWTH Aachen University under project \texttt{lect0148}.
\end{ack}

\medskip

{
\small
\bibliographystyle{unsrt}
\bibliography{references}
}

\appendix

\section{Training Curves}
\label{sec:training_curves}

\begin{figure}[!h]
  \centering
  \includegraphics[width=1\textwidth]{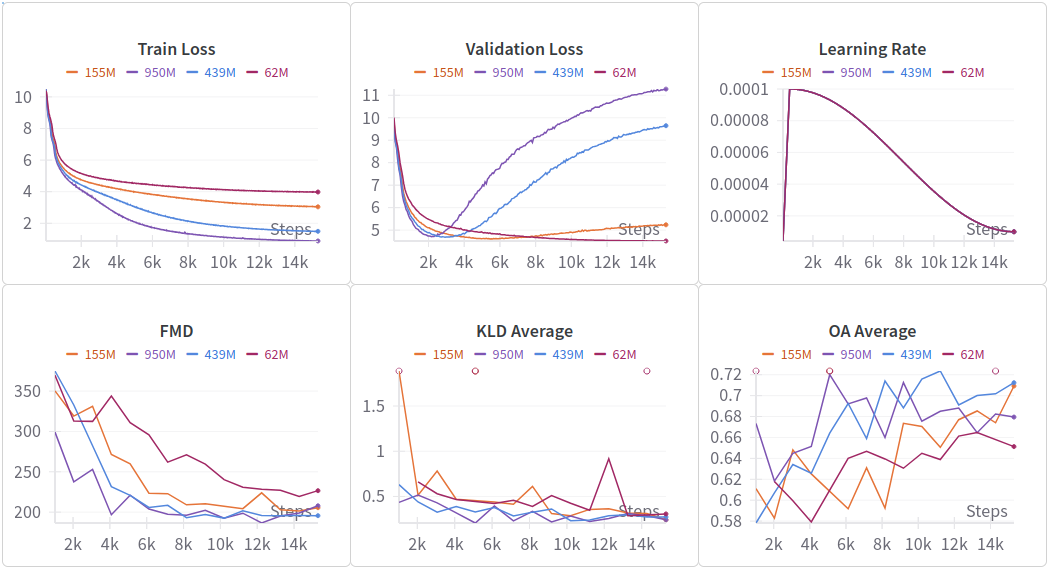}
  \caption{Training curves for the MAESTRO models of different sizes.}
  \label{fig:supp_curve_1}
\end{figure}

\FloatBarrier

\section{Musical Turing-like Test Analysis Results}

\begin{table}[!h]
  \centering
  \caption{Precision, Recall, and F1-Scores for the musical Turing-like test. ``Unsure'' responses were considered as incorrect.}
  \label{tab:turing_classification_report}
    \begin{tabular}{lccc}
        \toprule
        \textbf{Class} & \textbf{Precision} & \textbf{Recall} & \textbf{F1-score} \\
        \midrule
        Human     & \textbf{62.16\%} & \textbf{55.20\%} & \textbf{58.47\%} \\
        Generated & 60.19\% & 52.00\% & 55.79\% \\
        \bottomrule
    \end{tabular}
\end{table}

\FloatBarrier
\vspace*{0em}

\section{Full Evaluation Tables}
\definecolor{bandpurple}{RGB}{111,115,175}
\definecolor{bandgreen}{RGB}{197,224,180}
\definecolor{bandblue}{RGB}{189,215,238}
\definecolor{headergray}{RGB}{242,242,242}

\begin{table}[ht]
\centering
\small
\setlength{\tabcolsep}{5pt}
\renewcommand{\arraystretch}{0.9}
\caption{Subjective and objective evaluations of models trained on MAESTRO.}
\resizebox{0.95\textwidth}{!}{
\begin{tabular}{
    l
    *{4}{S[table-format=3.2]S[table-format=2.2]}
    S[table-format=3.2]S[table-format=2.2]
}
\toprule
\multicolumn{1}{c}{\textbf{Model}} &
\multicolumn{2}{c}{\textbf{62M}} &
\multicolumn{2}{c}{\textbf{155M}} &
\multicolumn{2}{c}{\textbf{439M}} &
\multicolumn{2}{c}{\textbf{950M}} &
\multicolumn{2}{c}{\textbf{MAESTRO}} \\
\midrule
\rowcolor{headergray}
\multicolumn{11}{l}{\textbf{Subjective Evaluation}}\\
Pleasingness        & \multicolumn{2}{c}{2.85} & \multicolumn{2}{c}{2.89} & \multicolumn{2}{c}{3.30} & \multicolumn{2}{c}{3.27} & \multicolumn{2}{c}{} \\
Authenticity        & \multicolumn{2}{c}{2.78} & \multicolumn{2}{c}{2.66} & \multicolumn{2}{c}{3.02} & \multicolumn{2}{c}{2.98} & \multicolumn{2}{c}{} \\
Novelty             & \multicolumn{2}{c}{2.89} & \multicolumn{2}{c}{3.03} & \multicolumn{2}{c}{3.35} & \multicolumn{2}{c}{3.27} & \multicolumn{2}{c}{} \\
\addlinespace[2pt]
\rowcolor{headergray}
\textbf{Average}    & \multicolumn{2}{c}{2.84} & \multicolumn{2}{c}{2.86} & \multicolumn{2}{c}{3.22} & \multicolumn{2}{c}{3.17} & \multicolumn{2}{c}{} \\
\addlinespace[2pt]
\rowcolor{bandpurple!25}
\multicolumn{11}{l}{\textbf{Absolute Evaluation}}\\
\multicolumn{1}{c}{} &
\multicolumn{1}{c}{\textbf{Mean}} & \multicolumn{1}{c}{\textbf{Std}} &
\multicolumn{1}{c}{\textbf{Mean}} & \multicolumn{1}{c}{\textbf{Std}} &
\multicolumn{1}{c}{\textbf{Mean}} & \multicolumn{1}{c}{\textbf{Std}} &
\multicolumn{1}{c}{\textbf{Mean}} & \multicolumn{1}{c}{\textbf{Std}} &
\multicolumn{1}{c}{\textbf{Mean}} & \multicolumn{1}{c}{\textbf{Std}} \\
\cmidrule(lr){2-3}\cmidrule(lr){4-5}\cmidrule(lr){6-7}\cmidrule(lr){8-9}\cmidrule(lr){10-11}
total\_used\_pitch  & {38.15} & {9.24} & {42.09} & {9.07} & {43.99} & {8.45} & {42.73} & {8.77} & {54.90} & {9.82} \\
pitch\_range        & {51.39} & {11.05} & {54.24} & {10.17} & {53.19} & {9.53} & {52.61} & {10.09} & {63.17} & {9.23} \\
avg\_pitch\_shift   & {9.36} & {4.25} & {9.88} & {3.51} & {11.05} & {3.31} & {10.67} & {3.82} & {12.35} & {2.79} \\
total\_used\_note   & {476.18} & {99.74} & {497.79} & {96.82} & {518.03} & {72.06} & {514.75} & {73.43} & {609.97} & {51.24} \\
avg\_IOI            & {0.15} & {0.10} & {0.13} & {0.09} & {0.14} & {0.10} & {0.14} & {0.09} & {0.09} & {0.05} \\
\addlinespace[2pt]
\rowcolor{headergray}
\textbf{Average Distance to MAESTRO}    & {0.33} & {10.48} & {0.25} & {9.61} & {0.23} & {4.61} & {0.25} & {5.04} & {0.00} & {0.00} \\
\addlinespace[2pt]
\rowcolor{bandgreen!35}
\multicolumn{11}{l}{\textbf{Relative Evaluation}}\\
\multicolumn{1}{c}{} &
\multicolumn{1}{c}{\textbf{KLD}} & \multicolumn{1}{c}{\textbf{OA}} &
\multicolumn{1}{c}{\textbf{KLD}} & \multicolumn{1}{c}{\textbf{OA}} &
\multicolumn{1}{c}{\textbf{KLD}} & \multicolumn{1}{c}{\textbf{OA}} &
\multicolumn{1}{c}{\textbf{KLD}} & \multicolumn{1}{c}{\textbf{OA}} &
\multicolumn{2}{c}{\textbf{}} \\
\cmidrule(lr){2-3}\cmidrule(lr){4-5}\cmidrule(lr){6-7}\cmidrule(lr){8-9}
total\_used\_pitch          & {0.03} & {71.63\%} & {0.03} & {81.79\%} & {0.01} & {86.46\%} & {0.03} & {82.64\%} & {} & {} \\
total\_pitch\_class\_hist   & {0.20} & {75.52\%} & {0.61} & {80.20\%} & {0.01} & {88.03\%} & {0.02} & {86.38\%} & {} & {} \\
pitch\_range                & {0.01} & {78.90\%} & {0.13} & {85.96\%} & {0.01} & {83.63\%} & {0.02} & {81.62\%} & {} & {} \\
avg\_pitch\_shift           & {0.05} & {76.91\%} & {0.00} & {86.25\%} & {0.00} & {93.18\%} & {0.01} & {87.74\%} & {} & {} \\
total\_used\_note           & {1.74} & {22.36\%} & {1.35} & {27.13\%} & {1.20} & {30.02\%} & {1.20} & {30.92\%} & {} & {} \\
avg\_IOI                    & {0.03} & {77.42\%} & {0.16} & {83.25\%} & {0.09} & {81.00\%} & {0.08} & {79.32\%} & {} & {} \\
note\_length\_hist          & {0.49} & {45.21\%} & {0.44} & {47.20\%} & {0.41} & {50.66\%} & {0.58} & {44.08\%} & {} & {} \\
note\_length\_transition\_matrix & {0.30} & {54.81\%} & {0.30} & {55.90\%} & {0.26} & {59.83\%} & {0.37} & {53.94\%} & {} & {} \\
\addlinespace[2pt]
\rowcolor{headergray}
\textbf{Average}         & {0.36} & {62.85\%} & {0.38} & {68.46\%} & {0.25} & {71.60\%} & {0.29} & {68.33\%} & {} & {} \\
\addlinespace[2pt]
\rowcolor{bandblue!35}
\textbf{FMD}                         & \multicolumn{2}{c}{210.01} & \multicolumn{2}{c}{187.53} & \multicolumn{2}{c}{176.40} & \multicolumn{2}{c}{176.38} & \multicolumn{2}{c}{} \\
\textbf{Perplexity}                  & \multicolumn{2}{c}{53.65} & \multicolumn{2}{c}{21.32} & \multicolumn{2}{c}{4.82} & \multicolumn{2}{c}{2.78} & \multicolumn{2}{c}{} \\
\bottomrule
\end{tabular}}
\label{tab:maestro_full}
\end{table}

\begin{table}[ht]
\centering
\small
\setlength{\tabcolsep}{5pt}
\renewcommand{\arraystretch}{0.9}
\caption{Subjective and objective evaluation of models pre-trained
on Aria-Deduped compared with the 155M MAESTRO model.}
\resizebox{0.82\textwidth}{!}{
\begin{tabular}{
    l
    *{4}{S[table-format=3.2]S[table-format=2.2]}
}
\toprule
\multicolumn{1}{c}{\textbf{Model}} &
\multicolumn{2}{c}{\textbf{155M}} &
\multicolumn{2}{c}{\textbf{155M-A-M}} &
\multicolumn{2}{c}{\textbf{155M-A-C}} &
\multicolumn{2}{c}{\textbf{MAESTRO}} \\
\midrule
\rowcolor{headergray}
\multicolumn{9}{l}{\textbf{Subjective Evaluation}}\\
Pleasingness        & \multicolumn{2}{c}{2.80} & \multicolumn{2}{c}{} & \multicolumn{2}{c}{3.25} & \multicolumn{2}{c}{} \\
Authenticity        & \multicolumn{2}{c}{2.80} & \multicolumn{2}{c}{} & \multicolumn{2}{c}{3.03} & \multicolumn{2}{c}{}\\
Novelty             & \multicolumn{2}{c}{2.95} & \multicolumn{2}{c}{} & \multicolumn{2}{c}{3.25} & \multicolumn{2}{c}{}\\
\addlinespace[2pt]
\rowcolor{headergray}
\textbf{Average}    & \multicolumn{2}{c}{2.85} & \multicolumn{2}{c}{} & \multicolumn{2}{c}{3.18} & \multicolumn{2}{c}{} \\
\addlinespace[2pt]
\rowcolor{bandpurple!25}
\multicolumn{9}{l}{\textbf{Absolute Evaluation}}\\
\multicolumn{1}{c}{} &
\multicolumn{1}{c}{\textbf{Mean}} & \multicolumn{1}{c}{\textbf{Std}} &
\multicolumn{1}{c}{\textbf{Mean}} & \multicolumn{1}{c}{\textbf{Std}} &
\multicolumn{1}{c}{\textbf{Mean}} & \multicolumn{1}{c}{\textbf{Std}} &
\multicolumn{1}{c}{\textbf{Mean}} & \multicolumn{1}{c}{\textbf{Std}} \\
\cmidrule(lr){2-3}\cmidrule(lr){4-5}\cmidrule(lr){6-7}\cmidrule(lr){8-9}
total\_used\_pitch  & {42.09} & {9.07} & {30.11} & {8.64} & {32.39} & {10.67} & {54.90} & {9.82}\\
pitch\_range        & {54.24} & {10.17} & {48.93} & {11.54} & {50.34} & {12.01} & {63.17} & {9.23}\\
avg\_pitch\_shift   & {9.88} & {3.51} & {9.71} & {3.02} & {10.18} & {3.32} & {12.35} & {2.79}\\
total\_used\_note   & {497.79} & {96.82} & {436.02} & {96.08} & {431.72} & {130.27} & {609.97} & {51.24}\\
avg\_IOI            & {0.13} & {0.09} & {0.27} & {0.18} & {0.25} & {0.34} & {0.09} & {0.05}\\
\addlinespace[2pt]
\rowcolor{headergray}
\textbf{Average Distance to MAESTRO}    & {0.25} & {9.61} & {0.65} & {9.74} & {0.58} & {16.70} & {0.00} & {0.00}\\
\addlinespace[2pt]
\rowcolor{bandgreen!35}
\multicolumn{9}{l}{\textbf{Relative Evaluation}}\\
\multicolumn{1}{c}{} &
\multicolumn{1}{c}{\textbf{KLD}} & \multicolumn{1}{c}{\textbf{OA}} &
\multicolumn{1}{c}{\textbf{KLD}} & \multicolumn{1}{c}{\textbf{OA}} &
\multicolumn{1}{c}{\textbf{KLD}} & \multicolumn{1}{c}{\textbf{OA}} &
\multicolumn{1}{c}{} & \multicolumn{1}{c}{} \\
\cmidrule(lr){2-3}\cmidrule(lr){4-5}\cmidrule(lr){6-7}
total\_used\_pitch          & {0.03} & {81.79\%} & {0.66} & {52.09\%} & {0.53} & {56.25\%} & {} & {}\\
total\_pitch\_class\_hist   & {0.61} & {80.20\%} & {0.25} & {70.96\%} & {0.29} & {68.59\%} & {} & {} \\
pitch\_range                & {0.13} & {85.96\%} & {0.20} & {71.89\%} & {0.19} & {73.49\%} & {} & {} \\
avg\_pitch\_shift           & {0.00} & {86.25\%} & {0.06} & {85.88\%} & {0.03} & {89.43\%} & {} & {} \\
total\_used\_note           & {1.35} & {27.13\%} & {2.28} & {24.10\%} & {2.65} & {15.64\%} & {} & {} \\
avg\_IOI                    & {0.16} & {83.25\%} & {0.65} & {51.63\%} & {0.66} & {49.49\%} & {} & {} \\
note\_length\_hist          & {0.44} & {47.20\%} & {1.48} & {41.16\%} & {1.21} & {40.00\%} & {} & {} \\
note\_length\_transition\_matrix & {0.30 } & {55.90\%} & {0.97} & {50.29\%} & {0.87} & {49.21\%} & {} & {}\\
\addlinespace[2pt]
\rowcolor{headergray}
\textbf{Average}         & {0.38 } & {68.46\%} & {0.82} & {56.00\%} & {0.80} & {55.26\%} & {} & {} \\
\addlinespace[2pt]
\rowcolor{bandblue!35}
\textbf{FMD}                        & \multicolumn{2}{c}{187.53} & \multicolumn{2}{c}{292.46} & \multicolumn{2}{c}{249.84} & \multicolumn{2}{c}{} \\
\bottomrule
\end{tabular}}
\label{tab:aria_full}
\end{table}

\begin{table}[ht]
\centering
\small
\setlength{\tabcolsep}{5pt}
\renewcommand{\arraystretch}{0.9}
\caption{Subjective and objective evaluation of models pre-trained
on Aria-Deduped and fine-tuned on MAESTRO compared with
the 155M MAESTRO model.}
\resizebox{0.82\textwidth}{!}{
\begin{tabular}{
    l
    *{4}{S[table-format=3.2]S[table-format=2.2]}
}
\toprule
\multicolumn{1}{c}{\textbf{Model}} &
\multicolumn{2}{c}{\textbf{155M}} &
\multicolumn{2}{c}{\textbf{155M-F-P}} &
\multicolumn{2}{c}{\textbf{155M-F-F}} &
\multicolumn{2}{c}{\textbf{MAESTRO}} \\
\midrule
\rowcolor{headergray}
\multicolumn{9}{l}{\textbf{Subjective Evaluation}}\\
Pleasingness        & \multicolumn{2}{c}{2.80} & \multicolumn{2}{c}{3.13} & \multicolumn{2}{c}{3.30} & \multicolumn{2}{c}{} \\
Authenticity        & \multicolumn{2}{c}{2.80} & \multicolumn{2}{c}{2.95} & \multicolumn{2}{c}{3.10} & \multicolumn{2}{c}{}\\
Novelty             & \multicolumn{2}{c}{2.95} & \multicolumn{2}{c}{3.20} & \multicolumn{2}{c}{3.35} & \multicolumn{2}{c}{}\\
\addlinespace[2pt]
\rowcolor{headergray}
\textbf{Average}    & \multicolumn{2}{c}{2.85} & \multicolumn{2}{c}{3.09} & \multicolumn{2}{c}{3.25} & \multicolumn{2}{c}{} \\
\addlinespace[2pt]
\rowcolor{bandpurple!25}
\multicolumn{9}{l}{\textbf{Absolute Evaluation}}\\
\multicolumn{1}{c}{} &
\multicolumn{1}{c}{\textbf{Mean}} & \multicolumn{1}{c}{\textbf{Std}} &
\multicolumn{1}{c}{\textbf{Mean}} & \multicolumn{1}{c}{\textbf{Std}} &
\multicolumn{1}{c}{\textbf{Mean}} & \multicolumn{1}{c}{\textbf{Std}} &
\multicolumn{1}{c}{\textbf{Mean}} & \multicolumn{1}{c}{\textbf{Std}} \\
\cmidrule(lr){2-3}\cmidrule(lr){4-5}\cmidrule(lr){6-7}\cmidrule(lr){8-9}
total\_used\_pitch  & {42.09} & {9.07} & {41.78} & {8.89} & {41.41} & {8.08} & {54.90} & {9.82}\\
pitch\_range        & {54.24} & {10.17} & {55.63} & {9.71} & {56.61} & {10.09} & {63.17} & {9.23}\\
avg\_pitch\_shift   & {9.88} & {3.51} & {11.30} & {3.56} & {10.48} & {3.59} & {12.35} & {2.79}\\
total\_used\_note   & {497.79} & {96.82} & {514.01} & {77.59} & {521.21} & {57.95} & {609.97} & {51.24}\\
avg\_IOI            & {0.13} & {0.09} & {0.12} & {0.08} & {0.11} & {0.07} & {0.09} & {0.05}\\
\addlinespace[2pt]
\rowcolor{headergray}
\textbf{Average Distance to MAESTRO}    & {0.25} & {9.61} & {0.19
} & {5.71} & {0.18} & {2.03} & {0.00} & {0.00}\\
\addlinespace[2pt]
\rowcolor{bandgreen!35}
\multicolumn{9}{l}{\textbf{Relative Evaluation}}\\
\multicolumn{1}{c}{} &
\multicolumn{1}{c}{\textbf{KLD}} & \multicolumn{1}{c}{\textbf{OA}} &
\multicolumn{1}{c}{\textbf{KLD}} & \multicolumn{1}{c}{\textbf{OA}} &
\multicolumn{1}{c}{\textbf{KLD}} & \multicolumn{1}{c}{\textbf{OA}} &
\multicolumn{1}{c}{} & \multicolumn{1}{c}{} \\
\cmidrule(lr){2-3}\cmidrule(lr){4-5}\cmidrule(lr){6-7}
total\_used\_pitch          & {0.03} & {81.79\%} & {0.09} & {80.45\%} & {0.09} & {80.98\%} & {} & {}\\
total\_pitch\_class\_hist   & {0.61} & {80.20\%} & {0.14} & {78.01\%} & {0.17} & {75.66\%} & {} & {} \\
pitch\_range                & {0.13} & {85.96\%} & {0.02} & {88.78\%} & {0.02} & {89.58\%} & {} & {} \\
avg\_pitch\_shift           & {0.00} & {86.25\%} & {0.04} & {88.44\%} & {0.06} & {84.66\%} & {} & {} \\
total\_used\_note           & {1.35} & {27.13\%} & {1.30} & {30.39\%} & {1.12} & {32.05\%} & {} & {} \\
avg\_IOI                    & {0.16} & {83.25\%} & {0.02} & {90.27\%} & {0.03} & {90.51\%} & {} & {} \\
note\_length\_hist          & {0.44} & {47.20\%} & {0.62} & {52.51\%} & {0.50} & {59.45\%} & {} & {} \\
note\_length\_transition\_matrix & {0.30 } & {55.90\%} & {0.46} & {60.06\%} & {0.36} & {66.20\%} & {} & {}\\
\addlinespace[2pt]
\rowcolor{headergray}
\textbf{Average}         & {0.38 } & {68.46\%} & {0.34} & {70.93\%} & {0.30} & {72.39\%} & {} & {} \\
\addlinespace[2pt]
\rowcolor{bandblue!35}
\textbf{FMD}                        & \multicolumn{2}{c}{187.53} & \multicolumn{2}{c}{193.45} & \multicolumn{2}{c}{187.34} & \multicolumn{2}{c}{} \\
\bottomrule
\end{tabular}}
\label{tab:fine-tune_full}
\end{table}

\end{document}